\documentclass[preprint]{ptephy_v2}

\preprintnumber{arXiv:2305.09309} 
\usepackage{hyperref}

\usepackage{graphicx}
\usepackage{dcolumn}

\usepackage[utf8]{inputenc}
\usepackage[T1]{fontenc}
\usepackage{mathptmx}
\usepackage{etoolbox}
\usepackage{enumitem}
\setlist[enumerate,1]{label={(\roman*)\,}}

\usepackage{comment}
\usepackage{MnSymbol}
\usepackage{url}
\usepackage{bbm}

\newcommand{\bra}{\langle}
\newcommand{\ket}{\rangle}

\newcommand{\mbb}[1]{\mathbb{#1}}
\newcommand{\mcal}[1]{\mathcal{#1}}
\newcommand{\mrm}[1]{\mathrm{#1}}
\newcommand{\msf}[1]{\mathsf{#1}}

\newcommand{\mfrak}[1]{\mathfrak{#1}}

\newcommand{\mbbm}[1]{\mathbbm{#1}}

\newcommand{\Setminus}{\, \backslash \,}

\newcommand{\im}{\mathrm{i}}

\newcommand{\LS}{L^{{S}}_{\theta}} 
\newcommand{\LR}{L^{{R}}_{\theta}} 
\newcommand{\JS}{\mcal{J}^{S}_{\theta}} 
\newcommand{\JM}{\mcal{J}^{\msf{M}}_{\theta}}

\newcommand{\ran}{\operatorname{ran}}
\renewcommand{\Re}{\operatorname{Re}}
\renewcommand{\Im}{\operatorname{Im}}

\newcommand{\Tr}{\operatorname{Tr}}

\newcommand{\tit}[1]{\textit{#1}}

\newcommand{\supp}{\operatorname{supp}} 
\newcommand{\sa}{\mathrm{sa}} 
\newcommand{\PM}{P^{\msf{M}}_{\theta}}
\newcommand{\DM}{\mcal{D}^{\msf{M}}_{\theta}}
\newcommand{\DS}{\mcal{D}^S_{\theta}}
\newcommand{\lM}{l^{\msf{M}}_{\theta}}

\newcommand{\st}{~\text{s.t.}~}
\usepackage{amsthm}
\newtheorem{definition}{Definition}[section] 
\newtheorem{theorem}[definition]{Theorem} 

\newtheorem{lemma}[definition]{Lemma}

\newcommand{\numberthis}{\addtocounter{equation}{1}\tag{\theequation}} 

\begin{document}


\title{Extension of the Watanabe--Sagawa--Ueda uncertainty relations for measurement errors to infinite-dimensional systems} 



\author{Ryosuke Nogami\thanks{Email: nogami.ryosuke.v3@s.mail.nagoya-u.ac.jp}}

\affil{Department of Complex Systems Science, Graduate School of Informatics,\\ Nagoya University\\Nagoya 464-8601, Japan}


\date{\today}

\begin{abstract}
    We extend the Watanabe--Sagawa--Ueda (WSU) uncertainty relations for measurement errors to infinite-dimensional systems. 
    The original WSU formulation provided a definition of measurement errors with a clear physical interpretation based on quantum estimation theory, but was restricted to finite-dimensional systems, excluding important observables such as position and momentum. Using pseudo-inverse forms of positive-semidefinite forms, we develop a framework for classical and quantum estimation theory for models whose parameter space is the set of full-rank states on an infinite-dimensional Hilbert space, and derive classical and quantum Cram\'{e}r--Rao inequalities. We extend the WSU measurement errors to both bounded and unbounded operators, and derive corresponding error-error uncertainty relations. 
    The resulting uncertainty relation inequalities are stronger than the original WSU bound due to an improved derivation method. Our results provide a theoretical framework for applying estimation-based uncertainty relations to observables with continuous values in infinite-dimensional systems.
\end{abstract}

\maketitle

\section{Introduction \label{sec:introduction}}

In quantum systems, there exist trade-off relationships between different types of information that an observer can know, which are collectively referred to as uncertainty relations. Heisenberg \cite{heisenberg1927} first proposed the uncertainty relation based on several thought experiments. 
For example, he considered measuring the position of an electron by irradiating it with gamma rays and deduced that the measurement accuracy of the electron's position improves as the wavelength of the gamma rays becomes shorter, but at the same time, the momentum change of the electron becomes larger.
Kennard \cite{1927Kennard} mathematically formulated the uncertainty relation as the product of the standard deviations of position and momentum being bounded below by $\hbar/2$, and Robertson \cite{1929Robertson} generalized it to any pair of observables:
\begin{equation}
    \sigma(A) \sigma(B) \ge \frac{1}{2} \big| \langle [A,B] \rangle \big|,
    \label{eq:KR UR}
\end{equation}
where $\sigma(X)$ denotes the standard deviation of an observable $X$ and $[A,B] := AB -BA$ is the commutator. This relation also holds for unbounded operators \cite{kraus1973expectation}.

The Kennard--Robertson relation \eqref{eq:KR UR} became widely recognized as the mathematical expression of the uncertainty relation. However, it does not address the measurement error or the disturbance caused by a measurement, which were central to Heisenberg's original thought experiments. The quantities $\sigma(A)$ and $\sigma(B)$ in \eqref{eq:KR UR} represent the standard deviations of separate, error-free measurements of physical quantities $A$ and $B$. Therefore, \eqref{eq:KR UR} is a statement about the properties of quantum states rather than about the properties of measurements.

The mathematical formulation of uncertainty relations for measurement errors became possible through the development of quantum measurement theory, with a pioneering contribution being the Arthurs--Kelly--Goodman relation \cite{ArthursKelly, Arthurs-Goodman1988}. However, their relation was restricted to unbiased measurements. Ozawa approached this problem differently, deriving the relationship between measurement error and disturbance to physical quantities in arbitrary measurements that may not satisfy the constraint of unbiasedness \cite{Ozawa2003}, as well as the relationship between errors in simultaneous measurements of two observables \cite{ozawa2004uncertainty}. Despite the generality of Ozawa's definitions of measurement error and disturbance, they suffered from a disadvantage: their physical interpretation was difficult, and consequently, they were not universally accepted.

In response to these challenges, Watanabe, Sagawa, and Ueda (WSU) defined measurement error and disturbance from the perspective of quantum estimation theory, and derived uncertainty relation inequalities that these quantities must satisfy \cite{watanabe2011uncertainty, watanabe2013formulation}. 
Their formulation offered two significant advantages: no unbiasedness constraint need to be imposed on the measurement itself, and the physical interpretation of the measurement errors and disturbances was clear. 
However, their definition also had limitations, as it was based on the theory of finite-parameter estimation and was therefore restricted to finite-dimensional systems whose states can be parametrized by a finite number of parameters. Position and momentum are important examples of physical quantities that are represented in infinite-dimensional systems. Excluding such systems makes it difficult to argue that the theory possesses sufficient generality as a physical theory.
To apply estimation-based measurement errors to infinite-dimensional systems, we need to consider appropriate sets of operators \cite{Holevo2011Probabilistic} that can be measured or estimated.
Moreover, although quantum estimation theory for infinite-dimensional systems with finite-parameter models has been discussed \cite{Holevo2011Probabilistic}, we must address infinite-parameter models to extend the WSU uncertainty relations.

In this paper, we extend the WSU's estimation-based measurement errors to infinite-dimensional systems and demonstrate that error-error uncertainty relations hold for these errors. 
To achieve this, we consider classical and quantum estimation theory for models whose parameter space is the set of full-rank states on an infinite-dimensional Hilbert space, and derive classical and quantum Cram\'{e}r--Rao inequalities.

This paper is organized as follows. Section \ref{sec:pseudo-inverse forms} provides the definition of the pseudo-inverse form of a positive-semidefinite form, which is needed to define measurement errors.
Section \ref{sec:classical estimation theory} addresses classical estimation theory for models induced by quantum measurements on full-rank states, introduces the classical Fisher information form, and derives the Cram\'{e}r--Rao inequality. 
Section \ref{sec:quantum estimation theory} considers quantum estimation theory for full-rank state models, introduces the quantum Fisher information forms, and derives the quantum Cram\'{e}r--Rao inequalities.
In Section \ref{sec:WSU uncertainty relation}, we extend WSU's estimation-based measurement errors to infinite-dimensional systems and derive error-error uncertainty relations for these errors.
Section \ref{sec:discussion} discusses the implications, limitations, and future directions of our results.

\section{Pseudo-inverse forms \label{sec:pseudo-inverse forms}}

In finite-dimensional estimation theory, the pseudo-inverse \cite{moore1920reciprocal,penrose1955generalized} of the Fisher information matrix is commonly employed due to its direct connection to the Cram\'{e}r--Rao bound. In the following sections, we introduce the classical and quantum Fisher information \tit{forms} rather than matrices or operators, which facilitates the extension to infinite-dimensional systems. Therefore, we consider the pseudo-inverses of (positive-semidefinite) sesquilinear forms.

\subsection{Definition}

Let $V$ be a complex linear space, and consider a sesquilinear form $\mcal{J}:V \times V \to \mbb{C}$ that is positive-semidefinite, i.e., $\mcal{J}(x, x) \geq 0$ for all $x \in V$. 
The linear space $V$ is called the domain of the sesquilinear form $\mcal{J}$ and is denoted by $\mcal{D}(\mcal{J})$.
For each linear functional $\varphi: V \to \mbb{C}$, we define 
\begin{equation}
    \tilde{\mcal{J}} (\varphi) :=
    \sup_{x \in V \Setminus \mcal{N}_{\mcal{J}}} \frac{|\varphi(x)|^2}{ \mcal{J}(x, x)},
\end{equation}
where $\mcal{N}_{\mcal{J}}$ is the null space of $\mcal{J}$, i.e., 
\begin{equation}
    \mcal{N}_{\mcal{J}} := \{ x \in V \mid \mcal{J}(x, x) = 0 \}.
\end{equation}
We define the domain of the pseudo-inverse of $\mcal{J}$ as
\begin{equation}
    \mcal{D}(\mcal{J}^{+})
    := \{\psi \in V' \mid \tilde{\mcal{J}}(\psi) < \infty \},
\end{equation}
where $V'$ denotes the dual space of $V$.
\begin{lemma}
    There exists a unique sesquilinear form $\mcal{J}^{+}$ on $\mcal{D}(\mcal{J}^{+})$ such that $\tilde{\mcal{J}}(\varphi) = \mcal{J}^{+}(\varphi, \varphi)$ for all $\varphi \in \mcal{D}(\mcal{J}^{+})$.
\end{lemma}
\begin{proof}
    Let 
    \begin{equation}
        \mcal{H}_0 := V / \mcal{N}_{\mcal{J}} 
        = \{[x] \mid x \in V\},
    \end{equation}
    where $[x] := x + \mcal{N}_{\mcal{J}}$ denotes the equivalence class to which $x \in V$ belongs.
    Then, we define a positive-definite inner product $(\cdot, \cdot)_{\mcal{J}}$ on the quotient space $\mcal{H}_0$ as
    \begin{equation}
        ([x], [y])_{\mcal{J}} := \mcal{J}(x, y), \quad \forall [x], [y] \in \mcal{H}_0.
    \end{equation}
    This definition is well-defined: if $n, m \in \mcal{N}_{\mcal{J}}$, then
    \begin{equation}
        \mcal{J}(x + n, y + m) = \mcal{J}(x, y) + \mcal{J}(n, y) + \mcal{J}(x, m) + \mcal{J}(n,m)
        = \mcal{J}(x, y),
    \end{equation}
    since $\mcal{J}(x, n) = 0$ holds for all $x \in V$ and $n \in \mcal{N}_{\mcal{J}}$.\footnote{Indeed, we can show the Cauchy--Schwarz inequality for the sesquilinear form $\mcal{J}$, i.e.,
    \begin{equation}
        |\mcal{J}(x, y)|^2 \leq \mcal{J}(x, x) \mcal{J}(y, y), \quad \forall x, y \in V.
    \end{equation}
    This implies that $\mcal{J}(x, n) = 0$ for all $x \in V$ and $n \in \mcal{N}_{\mcal{J}}$.}
    Thus, we can define the completion $\mcal{H}_{\mcal{J}}$ of $\mcal{H}_0$ with respect to the norm $\|x \|_{\mcal{J}} := \sqrt{(x, x)_{\mcal{J}}}$, in which the inner product $(\cdot, \cdot)_{\mcal{J}}$ is continuously extended.

    Each $\varphi \in \mcal{D}(\mcal{J}^{+})$ is bounded with respect to the seminorm $x \mapsto \sqrt{\mcal{J}(x, x)}$. We can therefore define a bounded linear functional $\tilde{\varphi}$ on $\mcal{H}_{0}$ by
    \begin{equation}
        \tilde{\varphi}([x]) := \varphi(x), \quad \forall [x] \in \mcal{H}_0.
    \end{equation}
    This is well-defined: for any $n \in \mcal{N}_{\mcal{J}}$, we have $\varphi(x + n) = \varphi(x) + \varphi(n) = \varphi(x)$ for all $x \in V$, since $|\varphi(n)| \leq M \sqrt{\mcal{J}(n, n)}=0$ for some $M > 0$. The linear functional $\tilde{\varphi}$ uniquely extends to a bounded linear functional on $\mcal{H}_{\mcal{J}}$.  
    By the Riesz representation theorem, there exists an element $y_{\varphi} \in \mcal{H}_{\mcal{J}}$ such that
    \begin{equation}
        \tilde{\varphi}(\xi) = (y_{\varphi}, \xi)_{\mcal{J}}, \quad \forall \xi \in \mcal{H}_{\mcal{J}}.
    \end{equation}
    Therefore,
    \begin{align*}
        \tilde{\mcal{J}}(\varphi) &= \sup_{x \in V \Setminus \mcal{N}_{\mcal{J}}} \frac{|\varphi(x)|^2}{\mcal{J}(x, x)} \\
        &= \sup_{\xi \in \mcal{H}_0: \xi \neq 0 } \frac{|(y_{\varphi}, \xi)_{\mcal{J}}|^2}{(\xi, \xi)_{\mcal{J}}} \\
        &= (y_{\varphi}, y_{\varphi})_{\mcal{J}}.
        \numberthis
    \end{align*}

    For $\varphi, \psi \in \mcal{D}(\mcal{J}^{+})$, we define
    \begin{equation}
        \mcal{J}^{+}(\varphi, \psi) := (y_{\psi}, y_{\varphi})_{\mcal{J}}.
    \end{equation}
    Then, we have $\mcal{J}^{+}(\varphi, \varphi) = \tilde{\mcal{J}}(\varphi) \geq 0$ for all $\varphi \in \mcal{D}(\mcal{J}^{+})$, and $\mcal{J}^{+}$ is sesquilinear because the mapping $\mcal{D}(\mcal{J}^{+}) \ni \varphi \mapsto y_{\varphi} \in \mcal{H}_{\mcal{J}}$ is antilinear. 

    The uniqueness follows from the polarization equality:
    \begin{equation}
        \mcal{J}^{+}(\varphi, \psi) = \frac{1}{4} \sum_{k=0}^{3} \im^k \mcal{J}^{+}(\varphi + (-\im)^k \psi, \varphi + (-\im)^k \psi), \quad \forall \varphi, \psi \in \mcal{D}(\mcal{J}^{+}). 
    \end{equation}
\end{proof}

We refer to $\mcal{J}^{+}$ as the \tit{pseudo-inverse form} of $\mcal{J}$.
We can similarly define the pseudo-inverse of a real positive-semidefinite bilinear form.

For two non-zero positive-semidefinite sesquilinear forms $\mcal{J}_1$ and $\mcal{J}_2$, we write $\mcal{J}_1 \leq \mcal{J}_2$ if $\mcal{D}(\mcal{J}_1) \supseteq \mcal{D}(\mcal{J}_2)$ and $\mcal{J}_1(\xi, \xi) \leq \mcal{J}_2(\xi, \xi)$ for all $\xi \in \mcal{D}(\mcal{J}_2)$. 
\begin{theorem}
    If $\mcal{J}_1 \leq \mcal{J}_2$, then $\mcal{J}_2^{+} \leq \mcal{J}_1^{+}$ holds. 
\end{theorem} 
\begin{proof}
    For each $\psi \in \mcal{D}(\mcal{J}_1^{+})$, we have
    \begin{align*}
        \mcal{J}_1^{+}(\psi, \psi) &= \sup_{\xi \in \mcal{D}(\mcal{J}_1) : \xi \notin \mcal{N}_{\mcal{J}_1}} \frac{|\psi(\xi)|^2}{\mcal{J}_1(\xi, \xi)} \\
        & \geq \sup_{\xi \in \mcal{D}(\mcal{J}_2) : \xi \notin \mcal{N}_{\mcal{J}_1}} \frac{|\psi(\xi)|^2}{\mcal{J}_1(\xi, \xi)} \\
        & \geq \sup_{\xi \in \mcal{D}(\mcal{J}_2) : \xi \notin \mcal{N}_{\mcal{J}_1}} \frac{|\psi(\xi)|^2}{\mcal{J}_2(\xi, \xi)} \\
        & \geq \sup_{\xi \in \mcal{D}(\mcal{J}_2) : \xi \notin \mcal{N}_{\mcal{J}_2}} \frac{|\psi(\xi)|^2}{\mcal{J}_2(\xi, \xi)} \\
        &= \tilde{\mcal{J}}_2(\psi).
        \numberthis
    \end{align*}
    This implies $\mcal{D}(\mcal{J}_1^{+}) \subseteq \mcal{D}(\mcal{J}_2^{+})$ and $\mcal{J}_2^{+}(\psi, \psi) \leq \mcal{J}_1^{+}(\psi, \psi)$ for all $\psi \in \mcal{D}(\mcal{J}_1^{+})$.
\end{proof}

\subsection{Justification for the pseudo-inverse form definition}\label{subsec:justification for pseudo-inverse form definition}
Now, we suppose that $V=\mcal{H}$ is a real or complex Hilbert space equipped with an inner product $\bra \cdot, \cdot \ket$. For each positive-semidefinite bounded operator $A$ on $\mcal{H}$, we can define a positive-semidefinite sesquilinear form $\mcal{J}_A: \mcal{H} \times \mcal{H} \to \mbb{C}$ by 
\begin{equation}
    \mcal{J}_A(x, y) := \bra x, A y \ket, \quad \forall x, y \in \mcal{H}.
\end{equation}

Let $f \in \mcal{D}(\mcal{J}_A^{+})$ be a linear functional .
Since
\begin{equation}
    |f(x)|^2 \leq \tilde{\mcal{J}}_A(f) \bra x , A x \ket \leq \tilde{\mcal{J}}_A(f) \|A\| \| x \|^2, \quad \forall x \in \mcal{H},
\end{equation}
it follows from the Riesz representation theorem that there exists a unique $y_f \in \mcal{H}$ such that $f(x) = \bra y_f, x \ket$ for all $x \in \mcal{H}$. Therefore, the following inequality holds:
\begin{equation}
    | \bra y_f, x \ket |^2 \leq \tilde{\mcal{J}}_A(f) \bra x, A x \ket, \quad \forall x \in \mcal{H}.
    \label{eq:inequality for y_f}
\end{equation}

For each $x \in \ker(A)$, we have 
\begin{equation}
    | \bra y_f , x \ket |^2 \leq \tilde{\mcal{J}}_A(f) \bra x, A x \ket = 0,
\end{equation}
which implies that $y_f \in \supp(A) := \ker(A)^{\perp} = \overline{\ran(A)}$.
For each $\epsilon > 0$, let 
\begin{equation}
    x_\epsilon := \left( \int_{\lambda \geq \epsilon} \lambda^{-1} d \msf{E}^A(\lambda)\right) y_f,
\end{equation}
where $\msf{E}^A$ is the spectral measure of $A$.
Then, substituting $x_\epsilon$ for $x$ in \eqref{eq:inequality for y_f} yields 
\begin{equation}
    \left(\int_{\lambda \geq \epsilon} \lambda^{-1} d \msf{E}^{A}_{y_f, y_f} (\lambda)\right)^2
    \leq \tilde{\mcal{J}}_A(f) \int_{\lambda \geq \epsilon} d \msf{E}^{A}_{y_f, y_f} (\lambda),
\end{equation}
where for a spectral measure $\msf{E}$ and $x, y \in \mcal{H}$, $\msf{E}_{x, y}$ denotes the complex-valued or signed measure defined by $\msf{E}_{x, y}(\Delta) := \bra x, \msf{E}(\Delta) y \ket$ for all $\Delta \in \mfrak{B}(\mbb{R})$. Here, $\mfrak{B}$ denotes the Borel $\sigma$-algebra over $\mbb{R}$.
Therefore, we have
\begin{equation}
     \int_{\lambda \geq \epsilon} d \msf{E}^{A}_{y_f, y_f} (\lambda) \leq \tilde{\mcal{J}}_A (f).
\end{equation}
By taking the limit $\epsilon \to +0$, we obtain
\begin{equation}
    \int_{\lambda > 0} d \msf{E}^{A}_{y_f, y_f} (\lambda) \leq \tilde{\mcal{J}}_A(f) < \infty.
\end{equation}
This implies that $y_f \in \mcal{D}(A^{-1/2})$ and $\| A^{-1/2} y_f \|^2 \leq \tilde{\mcal{J}}_A(f)$, where $A^{-1/2}$ is the self-adjoint operator on $\supp(A)$ defined by
\begin{gather}
    \mcal{D}(A^{-1/2}) := \left\{ x \in \supp(A) \,\middle|\, \int_{\lambda>0} \lambda^{-1} d \msf{E}^A_{x,x} (\lambda) < \infty \right\}, \\
    \bra x, A^{-1/2} y \ket := \int_{\lambda > 0} \lambda^{-1/2} d \msf{E}^A_{x, y}(\lambda), \quad \forall x, y \in \mcal{D}(A^{-1/2}).
\end{gather}

Conversely, for each $y \in \mcal{D}(A^{-1/2})$, it follows from $y = A^{1/2} A^{-1/2} y$ that 
\begin{align*}
    | \bra y, x \ket |^2 &= \big| \bra A^{-1/2} y, A^{1/2} x \ket \big|^2 \\
    & \leq \| A^{-1/2} y \|^2 \| A^{1/2} x \|^2 \\
    &= \| A^{-1/2} y \|^2 \bra x, A x\ket
\end{align*}
for all $x \in \mcal{H}$. Thus, for a linear functional $f$ defined by $f(x) := \bra y, x \ket$, we have $\tilde{\mcal{J}}_A(f) \leq \| A^{-1/2} y \|^2 < \infty$, which implies that $f \in \mcal{D}(\mcal{J}_A^{+})$. 

Therefore, we have established a one-to-one correspondence between $\mcal{D}(\mcal{J}_A^{+})$ and $\mcal{D}(A^{-1/2})$:
\begin{equation}
    \mcal{D}(\mcal{J}_A^{+}) = \{ f \in \mcal{H}' \mid \exists y_f \in \mcal{D}(A^{-1/2}) \st f(x) = \bra y_f, x \ket, \forall x \in \mcal{H} \}.
\end{equation} 
Moreover, since we have $\tilde{\mcal{J}}_A(f) = \| A^{-1/2} y_f \|^2$, the pseudo-inverse form $\mcal{J}_A^{+}$ is given by
\begin{equation}
    \mcal{J}_A^{+}(f, g) = \bra A^{-1/2} y_g, A^{-1/2} y_f \ket, \quad \forall f, g \in \mcal{D}(\mcal{J}_A^{+}).
\end{equation}

This justifies the definition of the pseudo-inverse form. However, we cannot express the pseudo-inverse form as $\mcal{J}_A^{+}(f,g) = \bra y_f, A^{-1} y_g \ket$ since the condition $y \in \mcal{D}(A^{-1/2})$ does not necessarily imply 
\begin{equation}
    y \in \mcal{D}(A^{-1}) 
    := \left\{
        x \in \supp(A) \,\middle|\, \int_{\lambda > 0} \lambda^{-2} d \msf{E}^A_{x,x} (\lambda) < \infty
    \right\}.
\end{equation}


\section{Classical estimation theory for models induced by quantum measurements on full-rank states \label{sec:classical estimation theory}}

To formulate the WSU uncertainty relations in infinite-dimensional systems, we need classical estimation theory for models with infinite-dimensional parameter spaces. Since dealing with general infinite-dimensional models is difficult, we concentrate on models induced by quantum measurements on full-rank states.

\subsection{Full-rank models}

Let $\mcal{H}$ be a separable complex Hilbert space equipped with an inner product $\bra \cdot, \cdot \ket$ and the norm $\| \psi \| := \sqrt{\bra \psi, \psi \ket}$.
Let $\mcal{B}(\mcal{H})$ and $\mcal{T}(\mcal{H})$ denote the set of bounded linear operators and trace-class operators on $\mcal{H}$, respectively. The set $\mcal{T}(\mcal{H})$ is a Banach space with respect to the trace norm $\|\cdot \|_1 := \Tr[|\cdot|]$.

We take the parameter space $\Theta$ to be the set of all full-rank density operators on $\mcal{H}$, where a density operator $\rho$ is said to be \tit{full-rank} if it satisfies one of the following equivalent conditions:
\begin{enumerate}
    \item $\ker (\rho) = \{0\}$, 
    \item $\supp (\rho) := (\ker (\rho))^{\perp} = \mcal{H}$,
    \item $\Tr[\rho A^{*} A] = 0$ implies $A = 0$ for any $A \in \mcal{B}(\mcal{H})$,
    \item There exists an orthonormal basis $(\phi_j)_{j=1}^{\infty}$ of $\mcal{H}$ consisting of eigenvectors of $\rho$ with positive eigenvalues.
\end{enumerate}

We consider the model $(\PM)_{\theta \in \Theta}$ induced by a positive operator-valued measure (POVM) $(\Omega, \mfrak{B}, \msf{M})$ through $\PM(\Delta) := \Tr[\theta \msf{M}(\Delta)]$ for all $\Delta \in \mfrak{B}$ and $\theta \in \Theta$. Here, $\Omega$ is a set and $\mfrak{B}$ denotes a $\sigma$-algebra on $\Omega$.

\subsection{Fisher information form}
Now, we fix a full-rank density operator $\theta \in \Theta$ and write its spectral decomposition as
\begin{equation}
    \theta = \sum_{j=1}^{\infty} \lambda_{j} |\phi_{j}\ket \bra \phi_j|,
    \label{eq:spectral decomposition of theta}
\end{equation}
where the $\lambda_j$ are positive eigenvalues of $\theta$ and $(\phi_j)_{j=1}^{\infty}$ is an orthonormal basis of $\mcal{H}$.
As the difference $\rho_1 - \rho_2$ between two density operators $\rho_1, \rho_2 \in \Theta$ is a traceless self-adjoint operator, it is natural to take the tangent space as a subset of the set of all traceless self-adjoint operators on $\mcal{H}$. Therefoure, we define the following two sets:
\begin{gather}
    \DM := \{\xi \in \mcal{T}_0(\mcal{H})_{\sa}  \mid \exists l^{\msf{M}}_\theta(\xi) \in L^2(\PM) \st \forall f \in L^{\infty}(\PM),\, E^{\msf{M}}_{\xi}[f] = E^{\msf{M}}_{\theta}[ l^{\msf{M}}_\theta(\xi) f] \}, 
    \\
    \mcal{D}_{0,\theta} := \mcal{T}_0(\mcal{H})_{\sa}
     \cap \operatorname{span} ( \{ |\phi_j \ket \bra \phi_k |\}_{j,k=1}^{\infty}),
\end{gather}
where 
\begin{gather}
    \mcal{T}_0(\mcal{H}) := \{ \xi \in \mcal{T}(\mcal{H}) \mid \Tr[\xi] = 0\} , \\
    \mcal{T}_0(\mcal{H})_{\sa} := \{ \xi \in \mcal{T}_0(\mcal{H}) \mid \xi = \xi^{*} \}, \\
    E^{\msf{M}}_{\xi}[f] := \int_{\Omega} f(\omega) dP^{\msf{M}}_{\xi}(\omega), \\
    P^{\msf{M}}_{\xi}(\Delta) := \Tr[\xi \msf{M}(\Delta)], \quad \forall \Delta \in \mfrak{B}. 
\end{gather}
The inclusion relation $\mcal{D}_{0,\theta} \subseteq \DM$ holds. Indeed, we will show later that $\mcal{D}_{0,\theta} \subseteq \mcal{D}^{S}_{\theta}$ (Lemma \ref{lem:D_0theta subset DS}) and $\DS \subseteq \DM$ (Theorem \ref{thm:quantum CR}), where $\DS$ is the domain of the SLD Fisher information form.

We refer to the function $l^{\msf{M}}_\theta(\xi)$ as the \tit{logarithmic derivative} of $\PM$ with respect to $\xi$.
Note that the relation $E^{\msf{M}}_{\xi}[f] = E^{\msf{M}}_{\theta}[ l^{\msf{M}}_\theta(\xi) f]$ for all $f \in L^{\infty}(\PM)$ uniquely determines $l^{\msf{M}}_\theta(\xi)$ up to $\PM$-almost everywhere equivalence. 
To see this, suppose that both $l_\theta(\xi)$ and $l'_\theta(\xi)$ satisfy the relation. For any $\Delta \in \mfrak{B}$, taking $f = 1_{\Delta}$ (the indicator function of $\Delta$) gives us $\int_{\Delta} l_\theta(\xi) d\PM = P^{\msf{M}}_{\xi}(\Delta) = \int_{\Delta} l'_{\theta}(\xi) d\PM$. Therefore, $\int_{\Delta} [l_\theta(\xi) - l'_{\theta}(\xi)] dP^{\msf{M}}_\theta = 0$ for all $\Delta \in \mfrak{B}$. We conclude that $l_\theta(\xi) = l'_\theta(\xi)$ $\PM$-almost everywhere, because if an integrable function has zero integral over every measurable set, then it vanishes $\PM$-almost everywhere.

Moreover, the mapping $l^{\msf{M}}_{\theta}: \DM \to L^2(\PM)$ is a closed linear operator. Indeed, if $(\xi_n)_{n=1}^{\infty}$ is a sequence in $\DM$ such that $\xi_n \to \xi \in \mcal{T}_0(\mcal{H})$ in the trace norm and $l^{\msf{M}}_{\theta}(\xi_n)$ converges to some $l \in L^2(\PM)$, then for each $f \in L^{\infty}(\PM)$, we have 
\begin{align*}
    \Tr\left[ \xi \int_{\Omega} f(\omega) d \msf{M}(\omega) \right]
    &= \lim_{n \to \infty} \Tr\left[ \xi_n \int_{\Omega} f(\omega) d \msf{M}(\omega) \right] \\
    &= \lim_{n \to \infty} E^{\msf{M}}_{\theta}[l^{\msf{M}}_{\theta}(\xi_n) f] \\
    &= E^{\msf{M}}_{\theta}[l f].
    \numberthis
\end{align*}
This implies that $\xi \in \DM$ and $l^{\msf{M}}_{\theta}(\xi) = l$. Therefore, $l^{\msf{M}}_{\theta}$ is closed.

We now make a brief remark on the relationship between our definition of the logarithmic derivative and the one given in Ref. \cite{nadaraya2012cramer}. 
In Ref. \cite{nadaraya2012cramer}, the logarithmic derivative is defined as the Radon--Nikod\'{y}m derivative of the signed measure $d_\theta \PM (\cdot)[\xi]$ with respect to the measure $\PM$, where $d_\theta \PM (\cdot)[\xi]$ denotes the derivative of the measure $\PM$ in the direction $\xi$ at $\theta$. If we define $l^{\msf{M}}_\theta(\xi)$ in this way, for any measurable function $f$ such that the integral $E^{\msf{M}}_{\xi}[f] = \int_{\Omega} f dP^{\msf{M}}_{\xi}$ exists, the relation $E^{\msf{M}}_{\xi}[f] = E^{\msf{M}}_{\theta}[l^{\msf{M}}_\theta(\xi) f]$ holds. However, this does not ensure that $l^{\msf{M}}_{\theta}(\xi)$ belongs to $L^2(\PM)$. 

We define a positive-semidefinite form $\mcal{J}^{\msf{M}}_\theta: \DM \times \DM \to \mbb{R}$ by
\begin{equation}
    \mcal{J}^{\msf{M}}_\theta(\xi, \eta) := E^{\msf{M}}_{\theta}[l^{\msf{M}}_\theta(\xi) l^{\msf{M}}_\theta(\eta)], \quad \forall \xi, \eta \in \DM.
    \label{eq:classical Fisher information form}
\end{equation}
We refer to $\mcal{J}^{\msf{M}}_\theta$ as the \tit{Fisher information form} of the model $(\PM)_{\theta \in \Theta}$ at $\theta$.

\subsection{Classical Cram\'{e}r--Rao inequality}

\begin{theorem}[Classical Cram\'{e}r--Rao inequality]
    For any function $f \in L^{\infty}(\PM)$, the inequality
    \begin{equation}
        \operatorname{Var}^{\msf{M}}_{\theta}[f] \geq (\mcal{J}^{\msf{M}}_\theta)^{+}( d_\theta \bar{f}, d_\theta \bar{f})
        \label{eq:classical Cramer-Rao inequality}
    \end{equation}
    holds. Here, $\operatorname{Var}^{\msf{M}}_{\theta}[\cdot]$ denotes the variance with respect to the measure $\PM$, $\bar{f}(\theta') := E^{\msf{M}}_{\theta'}[f]$, and $d_\theta \bar{f}[\xi] := E^{\msf{M}}_{\xi}[f]$ for any $\xi \in \mcal{T}_0(\mcal{H})$. 
\end{theorem}

\begin{proof}
    For each $\xi \in \mcal{D}^{\msf{M}}_\theta$, we have
    \begin{align*}
        \big|d_\theta \bar{f}[\xi]\big|^2 &= \big|E^{\msf{M}}_{\xi}[f]\big|^2 \\
        &= \big| E^{\msf{M}}_{\xi}[ f - \bar{f}(\theta) 1_{\Omega}] \big|^2 \quad (\text{since} \,\, E^{\msf{M}}_{\xi}[1_{\Omega}] = 0) \\
        &= \big|E^{\msf{M}}_{\theta}[l^{\msf{M}}_{\theta}(\xi) (f - \bar{f}(\theta) 1_{\Omega})]\big|^2 \\
        &\leq E^{\msf{M}}_{\theta}[(f-\bar{f}(\theta)1_{\Omega})^2] E^{\msf{M}}_{\theta}[(l^{\msf{M}}_{\theta}(\xi))^2] \quad (\text{by the Cauchy--Schwarz inequality}) \\
        &= \operatorname{Var}^{\msf{M}}_{\theta}[f] \mcal{J}^{\msf{M}}_\theta(\xi, \xi).
    \end{align*}
    Therefore, when $\mcal{J}^{\msf{M}}_\theta(\xi, \xi) \neq 0$, we have
    \begin{equation}
        \operatorname{Var}^{\msf{M}}_{\theta}[f] \geq \frac{\big|d_\theta \bar{f}[\xi]\big|^2}{\mcal{J}^{\msf{M}}_\theta(\xi, \xi)}.
    \end{equation}
    By taking the supremum over all $\xi \in \mcal{D}^{\msf{M}}_\theta$ such that $\mcal{J}^{\msf{M}}_\theta(\xi, \xi) \neq 0$, we obtain $d_\theta \bar{f} \in \DM$ and \eqref{eq:classical Cramer-Rao inequality}.
\end{proof}

\section{Quantum estimation theory for full-rank models \label{sec:quantum estimation theory}}

We formulate the quantum estimation theory for the model $(\theta)_{\theta \in \Theta}$, where $\Theta$ is the set of all full-rank density operators on a Hilbert space $\mcal{H}$ as defined in the previous section.

\subsection{Spaces of operators}

In the following, we fix a full-rank density operator $\theta \in \Theta$ and write its spectral decomposition as \eqref{eq:spectral decomposition of theta}.

Following Sec. 2 of Ref. \cite{Holevo2011Probabilistic}, we define the space $\mcal{L}^2_h(\theta)$ of square-summable symmetric operators as follows. Let $\mcal{B}(\mcal{H})_{\sa}$ denote the set of all bounded self-adjoint operators on $\mcal{H}$. We define the SLD inner product and the SLD norm as
\begin{equation}
    \bra A, B \ket_{\theta}^S := \frac{1}{2}\Tr[\theta \{A,B\}], \quad 
    \| A \|_{S,\theta}:= \sqrt{\bra A, A \ket_{\theta}^S} = \sqrt{\Tr[\theta A^2]}, \quad \forall A, B \in \mcal{B}(\mcal{H})_{\sa},
\end{equation}
where $\{A,B\} := AB + BA$ denotes the anti-commutator.
The SLD inner product is positive-definite due to the full-rank assumption on $\theta$. Therefore, we can define the real Hilbert space $\mcal{L}^2_h(\theta)$ as the completion of $\mcal{B}(\mcal{H})_{\sa}$ with respect to the SLD norm $\| \cdot \|_{S,\theta}$. 
Each element $A \in \mcal{L}^2_h(\theta)$ can be identified with a square-summable symmetric operator \cite[Theorem 2.8.1]{Holevo2011Probabilistic}. Here, a symmetric operator $A$ is said to be \tit{square-summable} with respect to $\theta$ if the domain $\mcal{D}(A)$ of $A$ is included in $\operatorname{span}(\{\phi_j\}_{j=1}^{\infty})$ and 
\begin{equation}
    \sum_{j=1}^{\infty} \lambda_j \| A \phi_j \|^2 
    = \bra A, A \ket_{\theta}^S < \infty.
\end{equation}
For example, an unbounded self-adjoint operator $A$ on $\mcal{H}$ such that 
\begin{equation}
    \bra A^2 \ket_{\theta} := \int_{\mbb{R}} \lambda^2 d P^{\msf{E}^A}_{\theta}(\lambda) < \infty
\end{equation}
belongs to $\mcal{L}^2_h(\theta)$, where $\msf{E}^A$ denotes the spectral measure of $A$. Note that each element of $\mcal{L}^2_h(\theta)$ does not necessarily have a self-adjoint extension in $\mcal{H}$.

Similarly, we define the RLD inner product and the RLD norm as
\begin{equation}
    \bra A, B\ket^R_\theta := \Tr[\theta B A^*], \quad
    \| A \|_{R, \theta} := \sqrt{\bra A, A\ket^R_{\theta}} = \sqrt{\Tr[\theta AA^*]}, \quad \forall A, B \in \mcal{B}(\mcal{H}).
\end{equation}
Then, we can define the complex Hilbert space $\mcal{L}_{\mrm{RLD}}(\theta)$ as the completion of $\mcal{B}(\mcal{H})$ with respect to the RLD norm $\|\cdot\|_{R,\theta}$.

\subsection{Quantum Fisher information forms}
Now, following Sec. 6 of Ref. \cite{Holevo2011Probabilistic}, we define the Fisher information as follows. Let
\begin{equation}
    \DS
    := \{\xi \in \mcal{T}_0(\mcal{H})_{\sa} \mid  
    \exists L^S_\theta(\xi) \in \mcal{L}^2_h(\theta) \,\st \forall A \in \mcal{B}(\mcal{H})_{\sa}, \,  \Tr[\xi A] = \bra L^S_{\theta}(\xi), A \ket^S_{\theta} \}.
\end{equation}
For each $\xi \in \DS$, the mapping $\mcal{B}(\mcal{H})_{\sa} \ni A \mapsto \Tr[\xi A] =  \bra L^S_{\theta}(\xi), A \ket^S_{\theta} \in \mbb{R}$ is a bounded linear functional with respect to the SLD norm $\|\cdot\|_{S,\theta}$ by the Cauchy-Schwarz inequality, and therefore the operator $L^S_\theta (\xi)$ is unique by the Riesz representation theorem. The mapping $L^S_\theta : \DS \ni \xi \mapsto L^S_{\theta}(\xi) \in \mcal{L}^2_h(\theta)$ is linear. 
We refer to $L^S_{\theta}(\xi)$ as the \tit{symmetric logarithmic derivative (SLD) operator}. 
We also define the \tit{SLD Fisher information form} $\mcal{J}^S_{\theta} : \DS \times \DS \to \mbb{R}$ by
\begin{equation}
    \mcal{J}^S_{\theta} (\xi, \eta) := \bra L^S_{\theta}(\xi), L^S_{\theta}(\eta) \ket^S_{\theta}, \quad \forall \xi,\eta \in \DS.
\end{equation}

Similarly, the RLD Fisher information can be defined. 
Let 
\begin{equation}
    \mcal{D}^R_{\theta} := \{\xi \in \mcal{T}_0(\mcal{H}) \mid \exists L^R_{\theta}(\xi)\in \mcal{L}_{\mrm{RLD}}(\theta) \, \st \forall A \in \mcal{B}(\mcal{H}), \, \Tr[\xi A] = \bra A^{*}, L^R_{\theta}(\xi) \ket^R_{\theta} \}.
\end{equation}
For each $\xi \in \mcal{D}^R_{\theta}(\xi)$, the operator $L^R_{\theta} \in \mcal{L}_{\mrm{RLD}}(\theta)$ is unique and is referred to as the \tit{right logarithmic derivative (RLD) operator}.  
The mapping $L^R_{\theta} : \mcal{D}^R_{\theta} \ni \xi \mapsto L^R_{\theta}(\xi) \in \mcal{L}_{\mrm{RLD}}(\theta)$ is complex linear.
We define the \tit{RLD Fisher information form} $\mcal{J}^R_{\theta}:\mcal{D}^R_{\theta} \times \mcal{D}^R_{\theta} \to \mbb{C}$ by
\begin{equation}
    \mcal{J}^R_{\theta} (\xi, \eta) := \bra L^R_{\theta}(\xi), L^R_{\theta}(\eta) \ket^R_{\theta}, \quad \forall \xi, \eta \in \mcal{D}^R_{\theta}.
\end{equation}
We also define the \tit{real RLD Fisher information form} $\mcal{J}^R_{\theta, \mbb{R}}: \mcal{D}^R_{\theta, \sa} \times  \mcal{D}^R_{\theta, \sa} \to \mbb{R}$ by
\begin{gather}
     \mcal{D}^R_{\theta, \sa} := \{ \xi \in \mcal{D}^R_{\theta} \mid \xi = \xi^{*}\},  \\
     \mcal{J}^{R}_{\theta, \mbb{R}} (\xi, \eta) := \Re \mcal{J}^R_{\theta}(\xi, \eta), \quad \forall \xi, \eta \in  \mcal{D}^R_{\theta, \sa}.
\end{gather}

\subsection{Quantum Cram\'{e}r--Rao inequalities}

\begin{theorem}[Quantum Cram\'{e}r--Rao inequalities]\label{thm:quantum CR}
    \begin{equation}
        \mcal{J}^{\msf{M}}_{\theta} \leq \mcal{J}^S_{\theta} \leq \mcal{J}^R_{\theta, \mbb{R}}.
    \end{equation}
\end{theorem}
\begin{proof}
    Proof of $\mcal{J}^{\msf{M}}_{\theta} \leq \mcal{J}^S_{\theta}$: 
    We define a map $\Gamma_{\msf{M}}: L^{\infty}(\PM) \to \mcal{B}(\mcal{H})$ by
    \begin{equation}
        \Gamma_{\msf{M}} (f) := \int_{\Omega} f d\msf{M}.
    \end{equation}
    This is a unital positive map on $L^{\infty}(\PM)$. Since $L^{\infty}(\PM)$ is an abelian $C^{*}$-algebra, the positive map $\Gamma_{\msf{M}}$ is also completely positive (CP) \cite{stinespring1955positive}. Therefore, by the Schwarz inequality for a unital CP map \cite{choi1974schwarz}, we have
    \begin{equation}
        \Gamma_{\msf{M}}(|f|^2) \geq \Gamma_{\msf{M}} (f)^2, \quad \forall f \in L^{\infty}_{\mbb{R}}(\PM).
        \label{eq:Schwarz inequality for unital CP maps}
    \end{equation}
    Here, for each $0<p\leq \infty$, $L^{p}_{\mbb{R}}(\PM)$ denotes the set of real-valued functions in $L^{p}(\PM)$. Taking the expectation $\Tr[\theta \cdot]$ of \eqref{eq:Schwarz inequality for unital CP maps} yields
    \begin{equation}
        E^{\msf{M}}_{\theta}[|f|^2] \geq \Tr[\theta \Gamma_{\msf{M}}(f)^2] = \| \Gamma_{\msf{M}}(f) \|_{S,\theta}^2.
    \end{equation}
   Since the left-hand side is the $L^2$-norm in $L^2_{\mbb{R}}(P^{\msf{M}})$, by the bounded linear transformation (BLT) theorem (e.g. \cite[Theorem 1.9.1]{megginson1998introduction}), the bounded linear map $\Gamma_{\msf{M}}: L^{\infty}(\PM) \to \mcal{B}(\mcal{H})$ uniquely extends to a linear contraction $\gamma_{\msf{M}}: L^2_{\mbb{R}}(\PM) \to \mcal{L}^2_h(\theta)$ such that $\gamma_{\msf{M}}f = \Gamma_{\msf{M}}(f)$ for all $f \in L^{\infty}_{\mbb{R}}(\PM)$. 

    For any $\xi \in \DS$ and any $f \in L^{\infty}_{\mbb{R}}(\PM)$, we have
    \begin{align*}
        E^{\msf{M}}_{\xi} [f] &= \Tr[\xi \Gamma_{\msf{M}}(f)] \\
        &= \bra L^S_{\theta}(\xi), \gamma_{\msf{M}} f \ket^S_{\theta} \\
        &= \bra \gamma_{\msf{M}}^{*} L^{S}_{\theta}(\xi), f \ket^{\msf{M}}_{\theta},
        \numberthis
    \end{align*}
    where $\bra g, h \ket^{\msf{M}}_{\theta} := E^{\msf{M}}_{\theta}[\bar{g}h]$ for $g, h \in L^2(\PM)$ is the inner product on $L^2(\PM)$ and $\gamma_{\msf{M}}^{*}: \mcal{L}^2_h \to L^2_{\mbb{R}} (\PM)$ is the adjoint map of $\gamma_{\msf{M}}$. This implies $\xi \in \DM$ (and therefore $\DS \subseteq \DM$) and $l^{\msf{M}}_{\theta}(\xi) = \gamma_{\msf{M}}^{*} \LS (\xi)$ for all $\xi \in \DS$.

    For each $\xi \in \DS$, we have
    \begin{align*}
        \JS(\xi, \xi) &= \|\LS(\xi)\|_{S,\theta}^2 \\
        &\geq \|\gamma_{\msf{M}}^{*} \| \|\LS(\xi)\|_{S,\theta}^2 \quad (\text{since} ~ \|\gamma_{\msf{M}}\| = \|\gamma_{\msf{M}}^{*}\| \leq 1) \\
        &\geq \| \gamma_{\msf{M}}^{*} \LS(\xi) \|_{\msf{M},\theta}^2 \\
        &= \bra \lM (\xi), \lM(\xi) \ket^{\msf{M}}_{\theta} \\
        &= \JM (\xi, \xi),
    \end{align*}
    where $\|f\|_{\msf{M},\theta} := \sqrt{E^{\msf{M}}_{\theta}[|f|^2]}$ is the $L^2$-norm on $L^2(\PM)$. Therefore, we have shown that $\JM \leq \JS$. 

    Proof of $\JS \leq \mcal{J}^R_{\theta, \mbb{R}}$: 
    If we define a real inner product
    \begin{equation}
        \bra A, B \ket^R_{\theta,\mbb{R}} := \Re \bra A, B \ket^R_{\theta}, \quad \forall A, B \in \mcal{L}_{\mrm{RLD}}(\theta)
    \end{equation}
    on $\mcal{L}_{\mrm{RLD}}(\theta)$, the complex Hilbert space $\mcal{L}_{\mrm{RLD}}(\theta)$ can be regarded as a \tit{real} Hilbert space. As a real Hilbert space, $\mcal{L}_{\mrm{RLD}}(\theta)$ is denoted by $\mcal{L}_{\mrm{RLD}}(\theta)_{\mbb{R}}$. 

    Since
    \begin{equation}
        \| A \|_{S,\theta}^2 = \Tr[\theta A^2] = \|A\|_{R,\theta}^2
    \end{equation}
    holds for all $A \in \mcal{B}(\mcal{H})_{\sa}$, the identity map on $\mcal{B}(\mcal{H})_{\sa}$ uniquely extends to a real linear isometry $\mcal{V}: \mcal{L}^2_h(\theta) \to \mcal{L}_{\mrm{RLD}}(\theta)_{\mbb{R}}$ such that $\mcal{V}A = A$ for all $A \in \mcal{B}(\mcal{H})_{\sa}$ by the BLT theorem. 
    We identify $\mcal{L}^2_h(\theta)$ with the real closed subspace $\mcal{V}\mcal{L}^2_h(\theta)$ in $\mcal{L}_{\mrm{RLD}}(\theta)_{\mbb{R}}$. Then, the dual map $\mcal{P} := \mcal{V}^{*}: \mcal{L}_{\mrm{RLD}}(\theta)_{\mbb{R}} \to \mcal{L}^2_h(\theta)$ can be regarded as the orthogonal projection onto $\mcal{V}\mcal{L}^2_h(\theta)$. 

    For each $\xi \in \mcal{D}^R_{\theta,\mbb{R}}$ and each $A \in \mcal{B}(\mcal{H})_{\sa}$, we have
    \begin{align*}
        \Tr[\xi A] &= \Re \Tr[\xi A] \\
        &= \bra A, \LR(\xi) \ket^R_{\theta, \mbb{R}} \\
        &= \bra \mcal{P} A, \LR(\xi) \ket^R_{\theta, \mbb{R}} \quad (\text{since}~ A \in \mcal{B}(\mcal{H})_{\sa} \subseteq \mcal{L}^2_h(\theta)) \\
        &= \bra A , \mcal{P} \LR (\xi) \ket^R_{\theta, \mbb{R}} \\
        &= \bra A, \mcal{P} \LR (\xi) \ket^S_{\theta}, 
        \numberthis
    \end{align*}
    where the last equality holds because $\bra \cdot, \cdot \ket^R_{\theta,\mbb{R}}$ coincides with $\bra \cdot, \cdot \ket^S_{\theta}$ on $\mcal{L}^2_h(\theta)$. This implies that $\xi \in \DS$ and $\LS(\xi) = \mcal{P}\LR(\xi)$. 

    Therefore, $\mcal{D}^{R}_{\theta,\mbb{R}} \subseteq \DS$ holds, and for each $\xi \in \mcal{D}^R_{\theta,\mbb{R}}$, we have
    \begin{align*}
        \mcal{J}^R_{\theta, \mbb{R}} (\xi, \xi) &= \| \LR (\xi) \|_{R,\theta}^2 \\
        &\geq \| \mcal{P} \LR (\xi) \|_{R, \theta}^2 \\
        &= \| \LS (\xi) \|_{S, \theta}^2 \\
        &= \JS (\xi, \xi).
    \end{align*}
    We have shown that $\JS \leq \mcal{J}^R_{\theta,\mbb{R}}$.
\end{proof}

\begin{lemma}{\label{lem:S-norm of A equals E^A-norm of (gamma_E^A)^* A}}
    For each $\theta \in \Theta$ and each $A \in \mcal{B}(\mcal{H})_{\sa}$, 
    \begin{equation}
        \| A \|_{S,\theta} = \| \gamma_{\msf{E}^A}^{*} A \|_{\msf{E}^A, \theta}
        \label{eq:S-norm of A}
    \end{equation}
    holds, where $\msf{E}^A$ denotes the spectral measure of $A$. 
\end{lemma}
\begin{proof}
    The spectral decomposition of $A$ is written as 
    \begin{equation}
        A = \int_{\mbb{R}} \lambda d \msf{E}^{A} (\lambda)
        = \int_{\mbb{R}} f d \msf{E}^{A},
    \end{equation}
    where $f (\lambda) := \lambda 1_{[-\|A\|, \|A\|]} (\lambda) \in L^{\infty}(P^{\msf{E}^A}_{\theta})$. For each $g \in L^{\infty}_{\mbb{R}}(P^{\msf{E}^A}_{\theta})$, we have
    \begin{align*}
        \bra \gamma_{\msf{E}^A}^{*} A, g \ket^{\msf{E}^A}_{\theta}
        &= \bra A, \gamma_{\msf{E}^A} g \ket^S_{\theta} \\
        &= \frac{1}{2} \Tr \left[ \theta \left\{ A ,\int_{\mbb{R}} g d\msf{E}^A \right\}\right] \\
        &= \Tr \left[\theta \int_{\mbb{R}} fg d\msf{E}^A \right] \\
        &= \bra f, g \ket^{\msf{E}^A}_{\theta},
        \numberthis
    \end{align*}
    where $\{A, B\} := AB + BA$ for $A, B \in \mcal{B}(\mcal{H})$. This yields $\gamma_{\msf{E}^A}^{*} A = f$. 

    Therefore, we obtain
    \begin{align*}
        \| \gamma_{\msf{E}^A}^{*} A \|_{\msf{E}^A,\theta}^2
        &= E^{\msf{E}^A}_{\theta}[f^2] \\
        &= \int_{\mbb{R}} f^2 d P^{\msf{E}^A}_{\theta} \\
        &= \Tr\left[ \theta \int_{\mbb{R}} f^2 d \msf{E}^A\right] \\
        &= \Tr[\theta A^2] \\
        &= \| A \|_{S,\theta}^2.
        \numberthis
    \end{align*}
    Thus, \eqref{eq:S-norm of A} holds.
\end{proof}

\begin{theorem}
    For each $\theta \in \Theta$ and each $\xi \in \DS$, the following identity holds:
    \begin{equation}
        \sup_{\msf{M}} \JM (\xi, \xi) = \JS (\xi, \xi).
        \label{eq:sup_M JM = JS}
    \end{equation}
    Here, the supremum $\sup_{\msf{M}}$ is taken over all POVMs on the measurable space $(\mbb{R},\mfrak{B}(\mbb{R}))$.
\end{theorem}
\begin{proof}
    By the quantum Cram\'{e}r--Rao inequality $\JM \leq \JS$, we have the left-hand side $\leq$ the right-hand side in \eqref{eq:sup_M JM = JS}. 

    Since $\mcal{L}^2_h(\theta)$ is the completion of $\mcal{B}(\mcal{H})_{\sa}$ with respect to the norm $\|\cdot\|_{S,\theta}$, for each $\xi \in \DS$, there is a sequence in $(L_n)_{n \in \mbb{N}}$ in $\mcal{B}(\mcal{H})_{\sa}$ such that $\lim_{n\to\infty}\|L_n - \LS(\xi)\|_{S,\theta} = 0$.
    We can define the spectral measure $\msf{M}_n := \msf{E}^{L_n}$ of $L_n$ for each $n \in \mbb{N}$, since $L_n$ is self-adjoint.

    Therefore, we obtain
    \begin{align*}
        \sqrt{\mcal{J}^{\msf{M}_n}_{\theta} (\xi, \xi)}
        &= \| l^{\msf{M}_n}_{\theta} (\xi) \|_{\msf{M}_n,\theta} \\
        &= \| \gamma_{\msf{M}_n}^{*} \LS (\xi) \|_{\msf{M}_n, \theta} \\
        &\geq \| \gamma_{\msf{M}_n}^{*} L_n \|_{\msf{M}_n, \theta} - \| \gamma_{\msf{M}_n}^{*} ( \LS (\xi) - L_n) \|_{\msf{M}_n, \theta} \quad (\text{by the triangular inequality}) \\
        &\geq \|L_n\|_{S,\theta} - \| \LS(\xi) - L_n \|_{S,\theta} \quad (\text{by Lemma \ref{lem:S-norm of A equals E^A-norm of (gamma_E^A)^* A} and}~ \|\gamma_{\msf{M}_n}^{*}\| \leq 1 ) \\
        &\xrightarrow{n\to\infty} \| \LS (\xi) \|_{S,\theta} \\
        &= \sqrt{\JS (\xi, \xi)}.
    \end{align*}
    This implies that the left-hand side $\geq$ the right-hand side in \eqref{eq:sup_M JM = JS}.
\end{proof}

\section{Watanabe--Sagawa--Ueda uncertainty relation}
\label{sec:WSU uncertainty relation}

In this section, we extend the WSU's estimation-based measurement errors to infinite-dimensional systems and derive error-error uncertainty relation inequalities.

\subsection{Expectation values, correlations, and commutators}

For $A \in \mcal{L}_{\mrm{RLD}}(\theta)$, we define the expectation value of $A$ as 
\begin{equation}
    \bra A \ket_{\theta} := \bra \mbbm{1}, A \ket^R_{\theta},
\end{equation}
where $\mbbm{1}$ denotes the identity operator on $\mcal{H}$. This extends the expectation value $\Tr[\theta A]$ for $A \in \mcal{B}(\mcal{H})$. 
We define a sesquilinear form $C_{\theta}:\mcal{L}_{\mrm{RLD}}(\theta) \times \mcal{L}_{\mrm{RLD}}(\theta) \to \mbb{C}$ and a real bilinear form $C^S_{\theta}: \mcal{L}^2_h(\theta) \times \mcal{L}^2_h(\theta) \to \mbb{R}$ by
\begin{equation}
    C_{\theta} (A, B)  := \bra A - \bra A \ket_{\theta} \mbbm{1}, B - \bra B \ket_{\theta} \mbbm{1} \ket^R_{\theta}
    = \bra A, B\ket^R_{\theta} - \overline{\bra A \ket_{\theta}} \bra B \ket_{\theta}, \quad \forall A, B \in \mcal{L}_{\mrm{RLD}}(\theta),
\end{equation}
and
\begin{align*}
    C^S_{\theta}(A,B) :=  \bra A - \bra A \ket_{\theta} \mbbm{1}, B - \bra B \ket_{\theta} \mbbm{1} \ket^S_{\theta}
    = \bra A, B \ket^S_{\theta} - \bra A \ket_{\theta} \bra B \ket_{\theta}, \quad \forall A, B \in \mcal{L}^2_h(\theta).
\end{align*}
We define the expectation value of the commutator as
\begin{equation}
    \frac{\im}{2} \bra [A, B] \ket_{\theta} 
    := \Im \bra A, B \ket^R_{\theta} = \Im C_{\theta}(A, B),
    \quad \forall A,B \in \mcal{L}^2_h(\theta).
\end{equation}
This definition is valid since the inclusion $\mcal{L}^2_h (\theta) \subseteq \mcal{L}^2 (\theta) \subseteq \mcal{L}_{\mrm{RLD}}(\theta)$ holds \cite{Holevo2011Probabilistic}, where $\mcal{L}^2(\theta)$ is the completion of $\mcal{B}(\mcal{H})$ with respect to the inner product $\bra A, B \ket^S_{\theta} := \frac{1}{2} \Tr [\theta (A^{*} B + B A^{*})]$.

\subsection{Pseudo-inverse of the SLD Fisher information form}

    For $A \in \mcal{B}(\mcal{H})_{\sa}$ and $\xi \in \mcal{T}_0(\mcal{H})_{\sa}$, we define
    \begin{equation}
        d_\theta \bra A \ket_{\theta} [\xi] := \Tr[\xi A].
    \end{equation}
    For $A \in \mcal{L}^2_h(\theta) \Setminus \mcal{B}(\mcal{H})_{\sa}$ and $\xi \in \DS$, we define 
    \begin{equation}
        d_\theta \bra A \ket_{\theta} [\xi] := \bra A, \LS (\xi) \ket^S_{\theta}.
    \end{equation}
    In both cases, $d_\theta \bra A \ket_\theta [\xi] =  \bra A , \LS (\xi) \ket^S_{\theta}$ holds for $A\in \mcal{L}^2_h(\theta)$ and $\xi \in \DS$.

\begin{lemma}
    For $A\in \mcal{L}^2_h(\theta)$ and $\xi \in \DS$, the following equation holds:
    \begin{equation}
        (\JS)^{+} (d_\theta \bra A \ket_\theta, d_\theta\bra A \ket_\theta)
        = \bra A, A \ket^S_\theta - \bra A \ket^2_\theta 
        =: \sigma_\theta (A)^2.
    \end{equation}
\end{lemma}
\begin{proof}
    We have 
    \begin{align*}
        (\JS)^{+}(d_\theta \bra A\ket_\theta, d_\theta \bra A \ket_\theta)
        &= \sup_{\xi \in \DS: \, \JS(\xi,\xi)\neq 0} \frac{| d_\theta \bra A \ket_\theta [\xi] |^2}{\JS(\xi,\xi)} \\
        &= \sup_{\xi \in \DS: \, \LS(\xi) \neq 0} \frac{\big| \bra A, \LS(\xi) \ket^S_\theta \big|^2}{\bra \LS(\xi), \LS(\xi) \ket^S_{\theta}} \\
        &= \| \mcal{P}_0 A \|_{S,\theta}^2,
        \numberthis
    \end{align*}
    where $\mcal{P}_0$ is the orthogonal projection from $\mcal{L}^2_h(\theta)$ onto the SLD norm closure $\overline{\ran \LS}$ of
    \begin{equation}
        \ran \LS = \{ \LS (\xi) \mid \xi \in \DS\}.
    \end{equation}

    The identity 
    \begin{equation}
        \ran \LS = \overline{\ran \LS} = \mcal{L}^2_h(\theta)_0 
        := \{ L \in \mcal{L}^2_h(\theta) \mid \bra L, \mbbm{1} \ket^S_\theta = 0\}
        \label{eq:ran LS}
    \end{equation}
    holds. We can establish \eqref{eq:ran LS} as follows. For each $\xi \in \DS$, we have 
    \begin{equation}
        \bra \LS (\xi), \mbbm{1} \ket^S_{\theta} = \Tr[\xi] = 0,
    \end{equation}
    which implies $\ran \LS \subseteq \mcal{L}^2_h(\theta)_0$. Conversely, for each $L \in \mcal{L}^2_h(\theta)_0$, we define
    \begin{align*}
        \xi_L :=& \frac{1}{2}\{\theta,L\} 
        := \frac{1}{2} [\sqrt{\theta} (L\sqrt{\theta})^{*} + (L\sqrt{\theta})\sqrt{\theta}]\\
        =& \frac{1}{2}\sum_{j=1}^{\infty} \lambda_j (|\phi_j\ket \bra L \phi_j | + |L\phi_j \ket \bra \phi_j|)
        \in \mcal{T}_0(\mcal{H})_{\sa},
        \numberthis
    \end{align*}
    where we used the fact that $L \sqrt{\theta}$ is of Hilbert--Schmidt class for $L \in \mcal{L}^2_h(\theta)$ \cite{Holevo2011Probabilistic}.
    Then, for each $A \in \mcal{B}(\mcal{H})_{\sa}$, we have
    \begin{equation}
        \Tr[\xi_L A] = \frac{1}{2} \sum_{j=1}^{\infty} (\bra L\phi_j , A\phi_j \ket + \bra A\phi_j, L\phi_j \ket)
        = \bra L, A \ket^S_\theta,
    \end{equation}
    which implies that $\xi_L \in \DS$ and $L = \LS(\xi_L) \in \ran \LS$. Therefore, we obtain $\ran \LS = \mcal{L}^2_h(\theta)_0$.
    Combining this with the fact that $\mcal{L}^2_h(\theta)_0$ is a closed subspace yields $\ran \LS = \mcal{L}^2_h(\theta)_0 = \overline{\mcal{L}^2_h(\theta)_0} = \overline{\ran \LS}$.

    The relation \eqref{eq:ran LS} implies that $\mcal{P}_0$ is the orthogonal projection onto $\mcal{L}^2_h(\theta)_0$, which is the subspace of vectors that are orthogonal to the unit vector $\mbbm{1} \in \mcal{L}^2_h(\theta)$.
    Therefore, we have
    \begin{equation}
        \mcal{P}_0 A = A - \bra A, \mbbm{1} \ket^S_\theta \mbbm{1}
        = A - \bra A \ket_\theta \mbbm{1}, \quad \forall A \in \mcal{L}^2_h(\theta).
    \end{equation}
    Thus, the equation
    \begin{align*}
        (\JS)^{+}(d_\theta \bra A \ket_\theta, d_\theta \bra A \ket_\theta) &= \| \mcal{P}_0 A \|_{S,\theta}^2 \\
        &= \bra A - \bra A \ket_\theta \mbbm{1}, A - \bra A \ket_\theta \mbbm{1} \ket^S_\theta \\
        &= \sigma_\theta (A)^2
        \numberthis
        \label{eq:JS^+(dA,dA)=variance}
    \end{align*}
    holds.
\end{proof}

\subsection{Measurement errors}

For an observable $A \in \mcal{B}(\mcal{H})_{\sa}$ and a POVM $(\mbb{R},\mfrak{B}(\mbb{R}),\msf{M})$, we define a measurement error by
\begin{equation}
    \varepsilon(A;\theta,\msf{M}):= \widetilde{\JM}(d_\theta \bra A \ket_\theta) - (\JS)^{+}(d_\theta \bra A\ket_\theta, d_\theta \bra A \ket_\theta)
    = \widetilde{\JM}(d_\theta \bra A \ket_\theta) - \sigma_\theta(A)^2.
    \label{eq:definition of the error for bounded operators}
\end{equation}
For $A \in \mcal{L}^2_h(\theta) \Setminus \mcal{B}(\mcal{H})_{\sa}$, we cannot define a measurement error using \eqref{eq:definition of the error for bounded operators}, since the domain of the linear functional $d_\theta \bra A \ket_\theta$ is $\DS (\subsetneq \DM)$, while $\widetilde{\JM}$ is defined only for linear functionals on $\DM$. 

We therefore define the measurement error for a possibly unbounded operator $A \in  \mcal{L}^2_h(\theta)$ in the following manner. 
\begin{lemma}\label{lem:D_0theta subset DS}
    For $\theta \in \Theta$, 
    \begin{enumerate}
        \item $\mcal{D}_{0,\theta} \subseteq \DS$, and
        \item $\LS (\mcal{D}_{0,\theta})$ is dense in $\mcal{L}^2_h(\theta)_0$.
    \end{enumerate}
\end{lemma}
\begin{proof}
    (i): Let $\eta_{j,k} := |\phi_j\ket \bra \phi_k | + |\phi_k\ket\bra \phi_j|$ and $\gamma_{j,k} := \im (|\phi_j\ket \bra \phi_k|-|\phi_k\ket \bra \phi_j| )$ for $j,k \in \mbb{N}$. 
    For each $A\in \mcal{B}(\mcal{H})_{\sa}$, we have
    \begin{align*}
        \left\langle \frac{2}{\lambda_j + \lambda_k} \eta_{j,k}, A \right\rangle^S_\theta 
        &= \frac{1}{2} \Tr \left[\theta \left\{ \frac{2}{\lambda_j + \lambda_k} \eta_{j,k}, A \right\}\right] \\
        &= \sum_{l=1}^{\infty} \lambda_l \frac{ \delta_{l,j} \bra \phi_k, A \phi_l\ket + \delta_{l,k} \bra \phi_j, A \phi_l \ket + \bra \phi_l, A \phi_j \ket \delta_{k,l} + \bra \phi_l, A \phi_k \ket \delta_{j,l} }{\lambda_j + \lambda_k} \\
        &= \bra \phi_k , A \phi_j \ket + \bra \phi_j, A \phi_k \ket \\
        &= \Tr[\eta_{j,k} A].
        \numberthis
    \end{align*}
    Therefore, we obtain
    \begin{equation}
        \eta_{j,k} \in \DS, \quad \LS(\eta_{j,k}) = \frac{2}{\lambda_j + \lambda_k} \eta_{j,k}
        \label{eq: LS(eta_jk)}
    \end{equation}
    for $j \neq k$, and 
    \begin{equation}
        \sum_{j=1}^{n} c_{j} \eta_{j,j} \in \DS, \quad \LS\left(\sum_{j=1}^{n} c_{j} \eta_{j,j}\right)
        = \sum_{j=1}^{n} \frac{c_j}{\lambda_j} \eta_{j,j}
    \end{equation}
    for $j \in \mbb{N}$ and a sequence $(c_{j})_{j=1}^{n} \subseteq \mbb{R}$ such that $\sum_{j=1}^{n} c_{j}=0$. 
    In a similar manner, we can show that
    \begin{equation}
        \gamma_{j,k} \in \DS, \quad \LS(\gamma_{j,k}) = \frac{2}{\lambda_j + \lambda_k} \gamma_{j,k}
        \label{eq: LS(gamma_jk)}
    \end{equation}
    for $j \neq k$.

    For any $\xi \in \mcal{D}_{0,\theta}$, there exist coefficients $(c_{j,k})_{j,k=1}^{n} \subseteq \mbb{C}$ such that
    \begin{subequations}
        \begin{gather}
            \xi = \sum_{j,k=1}^{n} c_{j,k} |\phi_j\ket \bra \phi_k|,\\
            c_{j,k} = c_{j,k}^{*},\quad \forall j,k = 1,\ldots,n, \\
            \sum_{j=1}^{n} c_{j,j} = 0.
        \end{gather}
    \end{subequations}
    Therefore, we have
    \begin{equation}
        \xi = \sum_{j=1}^n \frac{c_{j,j}}{2} \eta_{j,j} 
        + \sum_{1\leq j \neq k \leq n} \frac{c_{j,k}}{2} ( \eta_{j,k} - \im \gamma_{j,k}) \in \DS,
    \end{equation}
    which implies that $\mcal{D}_{0,\theta} \subseteq \DS$.

    (ii): The inclusion $\LS(\mcal{D}_{0,\theta}) \subseteq \mcal{L}^2_h(\theta)_0$ follows immediately from $\mcal{L}^2_h(\theta)_0 = \ran \LS = \LS (\DS)$.
    To show the subspace $\LS(\mcal{D}_{0, \theta})$ is dense in $\mcal{L}^2_h(\theta)_0$, it suffices to show that the only element $L \in \mcal{L}^2_h(\theta)_0$ that satisfies $\bra \LS(\xi), L \ket^S_{\theta} = 0$ for all $\xi \in \mcal{D}_{0,\theta}$ is $L=0$.
    Suppose that for $L\in \mcal{L}^2_h(\theta)_0$, we have $\bra \LS(\xi), L \ket^S_{\theta} = 0$ for all $\xi \in \mcal{D}_{0,\theta}$.
    Then, we obtain $\bra \LS(\eta_{j,k}), L \ket^S_{\theta} = \bra \LS(\gamma_{j,k}), L \ket^S_\theta = 0$ for $j \neq k$. Thus, $\bra \phi_j, L\phi_k \ket = 0$ for $j\neq k$. 
    For each $j\in \mbb{N}$, let
    \begin{equation}
        \omega_j := |\phi_j\ket \bra \phi_j| - |\phi_{j+1} \ket \bra \phi_{j+1}| \in \mcal{D}_{0,\theta}.
    \end{equation}
    Since 
    \begin{equation}
        0 = \bra \LS(\omega_j) , L\ket^S_{\theta} = \bra \phi_j, L\phi_j \ket - \bra \phi_{j+1}, L \phi_{j+1}\ket
    \end{equation}
    for each $j \in \mbb{N}$, we have $\bra \phi_j, L \phi_k  \ket = c \delta_{j,k}$ for some $c \in \mbb{R}$, i.e., $L=c \mbbm{1}$. Since $L \in \mcal{L}^2_h(\theta)_0$, we obtain $c=0$, and hence $L=0$.
\end{proof}

Let $\mcal{J}^{\msf{M}}_{0,\theta} : \mcal{D}_{0,\theta} \times \mcal{D}_{0,\theta} \to \mbb{R}$ be the restriction of $\JM$ to $\mcal{D}_{0,\theta}$.
\begin{lemma}
    For $A \in \mcal{L}^2_h(\theta)$ and a POVM $(\mbb{R},\mfrak{B}(\mbb{R}),\msf{M})$, the following inequality holds:
    \begin{equation}
        \widetilde{\mcal{J}^{\msf{M}}_{0,\theta}}(d_\theta \bra A \ket_\theta) \geq \sigma_\theta(A)^2.
        \label{eq:pseudo-inverse of restricted FI form and variance}
    \end{equation}
\end{lemma}
\begin{proof}
    We need only consider the case where $\widetilde{\mcal{J}^{\msf{M}}_{0,\theta}}(d_\theta \bra A \ket_\theta)$ is finite. Then, we have
    \begin{align*}
        \widetilde{\mcal{J}^{\msf{M}}_{0,\theta}}(d_\theta \bra A \ket_\theta) 
        &= \sup_{\xi \in \mcal{D}_{0,\theta} : \, \JM(\xi, \xi) \neq 0} \frac{| d_\theta \bra A \ket_\theta [\xi] |^2}{\JM(\xi,\xi)} \\
        &\geq \sup_{\xi \in \mcal{D}_{0,\theta}:\, \LS(\xi) \neq 0} \frac{ \big| \bra \LS(\xi), A \ket^S_{\theta} \big|^2}{\bra \LS(\xi), \LS(\xi) \ket^S_\theta} \\
        &= \|\mcal{P}_0 A\|_{S,\theta}^2,
        \numberthis
    \end{align*}
    where the inequality follows from $\JM \leq \JS$, and the last equality follows from Lemma \ref{lem:D_0theta subset DS}.
    Therefore, inequality \eqref{eq:pseudo-inverse of restricted FI form and variance} follows from \eqref{eq:JS^+(dA,dA)=variance}.
\end{proof}

For $A \in \mcal{L}^2_h(\theta)$ and a POVM $\msf{M}$, we define another measurement error by
\begin{equation}
    \varepsilon_0(A; \theta, \msf{M})
    := \widetilde{\mcal{J}^{\msf{M}}_{0,\theta}} (d_\theta \bra A \ket_\theta) - \sigma_\theta(A)^2.
\end{equation} 
When $\mcal{H}$ is finite-dimensional, $\varepsilon(A; \theta, \msf{M})$ and $\varepsilon_0(A; \theta, \msf{M})$ coincide since $\DM$ and $\mcal{D}_{0,\theta}$ are identical, and hence so are $\JM$ and $\mcal{J}^{\msf{M}}_{0,\theta}$.

\begin{lemma}{\label{lem:expllicit formula for tilde JM0}}
    The explicit formula for $\widetilde{\mcal{J}^{\msf{M}}_{0,\theta}} (d_\theta \bra A \ket_\theta)$ is given by
    \begin{equation}
        \widetilde{\mcal{J}^{\msf{M}}_{0,\theta}} (d_\theta \bra A \ket_\theta)
        = \begin{cases}
            \| (\gamma_{\msf{M}} \gamma_{\msf{M}}^{*})^{-1/2} A' \|_{S,\theta}^2 & \text{if } A' \in \mcal{D}( (\gamma_{\msf{M}} \gamma_{\msf{M}}^{*})^{-1/2}); \\ 
            \infty & \text{otherwise},
        \end{cases}
    \end{equation}
    where $A' := \mcal{P}_0 A = A - \bra A \ket_\theta \mbbm{1}$.   
\end{lemma}
\begin{proof}
    Since $d_\theta \bra A \ket_\theta = d_\theta \bra A' \ket_{\theta}$, we have
    \begin{align*}
        \widetilde{\mcal{J}^{\msf{M}}_{0,\theta}} (d_\theta \bra A \ket_\theta) &= \widetilde{\mcal{J}^{\msf{M}}_{0,\theta}} (d_\theta \bra A' \ket_\theta) \\
        &= \sup_{\xi \in \mcal{D}_{0,\theta}: \, \JM(\xi,\xi) \neq 0} \frac{\big|d_\theta \bra A'\ket_\theta[\xi]\big|^2}{\JM(\xi,\xi)} \\
        &= \sup_{L \in \LS(\mcal{D}_{0,\theta}): \, \gamma_{\msf{M}}^{*}L \neq 0} \frac{\big| \bra L, A' \ket^S_{\theta}\big|^2}{\bra \gamma_{\msf{M}}^{*} L, \gamma_{\msf{M}}^{*} L \ket^{\msf{M}}_{\theta}} \\
        &= \sup_{L \in \mcal{L}^2_h(\theta)_0 : \, \bra L, \gamma_{\msf{M}}\gamma_{\msf{M}}^{*}L \ket^S_\theta \neq 0} \frac{\big| \bra L, A' \ket^S_{\theta}\big|^2}{\bra L, \gamma_{\msf{M}}\gamma_{\msf{M}}^{*} L \ket^{S}_{\theta}},
        \numberthis
        \label{eq:tilde JM0}
    \end{align*}
    where the third equality follows from $\lM(\xi) = \gamma_{\msf{M}}^{*} \LS(\xi)$, and the last equality follows from the density of $\LS(\mcal{D}_{0,\theta})$ in $\mcal{L}^2_h(\theta)_0$ (Lemma \ref{lem:D_0theta subset DS}).

    We define a bilinear functional $\Phi^{\msf{M}}_\theta : \mcal{L}^2_h(\theta)_0 \times \mcal{L}^2_h(\theta)_0 \to \mbb{R}$ by 
    \begin{equation}
        \Phi^{\msf{M}}_\theta (K,L) := \bra K, \gamma_{\msf{M}} \gamma_{\msf{M}}^{*} L \ket^S_{\theta},
        \quad \forall K,L \in \mcal{L}^2_h(\theta)_0,
    \end{equation}
    and define a linear functional $\varphi_A :\mcal{L}^2_h(\theta)_0 \to \mbb{R}$ by
    \begin{equation}
        \varphi_A (L) := \bra A', L \ket^S_\theta,
        \quad \forall L \in \mcal{L}^2_h(\theta)_0.
    \end{equation}
    Equation \eqref{eq:tilde JM0} implies
    \begin{equation}
        \widetilde{\mcal{J}^{\msf{M}}_{0,\theta}} (d_\theta \bra A \ket_\theta) = \widetilde{\Phi^{\msf{M}}_\theta}(\varphi_A).
    \end{equation}
    From the same argument as in Subsection \ref{subsec:justification for pseudo-inverse form definition}, we have
    \begin{equation}
         \widetilde{\Phi^{\msf{M}}_\theta}(\varphi_A) = \begin{cases}
             \| (\gamma_{\msf{M}} \gamma_{\msf{M}}^{*})^{-1/2} A' \|_{S,\theta}^2 & \text{if } A' \in \mcal{D}((\gamma_{\msf{M}} \gamma_{\msf{M}}^{*})^{-1/2}) ; \\
             \infty & \text{otherwise,}
         \end{cases}
    \end{equation}
    which completes the proof.
\end{proof}

\begin{theorem}
    The measurement errors for self-adjoint $A \in \mcal{L}^2_h(\theta)$ when measured using its own spectral measure are zero:
    \begin{equation}
        \varepsilon(A;\theta,\msf{E}^A) = 0 
    \end{equation}
    for $A \in \mcal{B}(\mcal{H})_{\sa}$, and
    \begin{equation}
        \varepsilon_0(A; \theta, \msf{E}^A) = 0
    \end{equation}
    for self-adjoint $A \in \mcal{L}^2_h(\theta)$.

    This property is referred to as the \tit{soundness} of the measurement errors \cite{ozawa2019soundness}.
\end{theorem}

\begin{proof}
    Case 1: When $A \in \mcal{B}(\mcal{H})_{\sa}$: Let $f(\lambda) := \lambda 1_{[-\|A\|,\|A\|]}(\lambda)$. Then, we have $f \in L^{\infty}(P^{\msf{E}^A}_{\theta})$ and $A = \int_{\mbb{R}} f d\msf{E}^A$.
    The expectation value of $f$ is given by
    \begin{equation}
        E^{\msf{E}^A}_{\theta}[f] = \int_{\mbb{R}} f(\lambda) dP^{\msf{E}^A}_{\theta}(\lambda) = \bra A \ket_{\theta},
    \end{equation}
    and the variance is given by
    \begin{equation}
        \operatorname{Var}^{\msf{E}^A}_{\theta}[f]
        = \int_{\mbb{R}} (f(\lambda) - \bra A \ket_\theta)^2 dP^{\msf{E}^A}_{\theta}
        = \sigma_\theta(A)^2.
    \end{equation}
    Therefore, it follows from \eqref{eq:classical Cramer-Rao inequality} that 
    \begin{equation}
        \sigma_\theta(A)^2 = \operatorname{Var}^{\msf{E}^A}_{\theta}[f]
        \geq (\mcal{J}^{\msf{E}^A}_{\theta})^{+}(d_\theta \bra A \ket_\theta, d_\theta \bra A \ket_\theta) \geq \sigma_\theta(A)^2,
    \end{equation}
    which implies that $\varepsilon(A;\theta,\msf{E}^A)=0$.

    Case 2: When $A \in \mcal{L}^2_h(\theta)$ is self-adjoint: 
    For each $n \in \mbb{N}$, we define
    \begin{equation}
        f_n (\lambda) := \lambda 1_{[-n,n]}(\lambda),
    \end{equation}
    and 
    \begin{equation}
        A_n := \int_{\mbb{R}} f_n d \msf{E}^A \in \mcal{B}(\mcal{H})_{\sa}.
    \end{equation}
    By the dominated convergence theorem, we obtain
    \begin{equation}
        \|A-A_n\|_{S,\theta}^2
        = \int_{\mbb{R}} (\lambda - f_n(\lambda))^2 dP^{\msf{E}^A}_\theta(\lambda) 
        \xrightarrow{n\to\infty} 0.
    \end{equation}
    Following a similar argument as in the proof of Lemma \ref{lem:S-norm of A equals E^A-norm of (gamma_E^A)^* A}, we can show that
    \begin{equation}
        f_n = \gamma_{\msf{E}^A}^{*} A_n,
    \end{equation}
    and hence 
    \begin{equation}
        A_n = \int_{\mbb{R}} f_n d\msf{E}^A
        = \gamma_{\msf{E}^A} f_n 
        = \gamma_{\msf{E}^A} \gamma_{\msf{E}^A}^{*} A_n.
    \end{equation}
    This implies that $A_n$ is an eigenvector of $\gamma_{\msf{E}^A} \gamma_{\msf{E}^A}^{*}$ with eigenvalue $1$. 
    Therefore, we have $A_n \in \mcal{D}((\gamma_{\msf{E}^A} \gamma_{\msf{E}^A}^{*})^{-1/2})$ and $A_n = (\gamma_{\msf{E}^A} \gamma_{\msf{E}^A}^{*})^{-1/2} A_n$.
    Since $(\gamma_{\msf{E}^A} \gamma_{\msf{E}^A}^{*})^{-1/2} A_n = A_n \to A$ in $\mcal{L}^2_h(\theta)$ as $n \to \infty$ and the self-adjoint operator $(\gamma_{\msf{E}^A} \gamma_{\msf{E}^A}^{*})^{-1/2}$ is closed, we have $A \in \mcal{D}((\gamma_{\msf{E}^A} \gamma_{\msf{E}^A}^{*})^{-1/2})$ and $A = (\gamma_{\msf{E}^A} \gamma_{\msf{E}^A}^{*})^{-1/2} A$.
    \footnote{This follows from the properties of the spectral measure of a self-adjoint operator: if $\psi$ is an eigenvector of a self-adjoint operator $T$ with eigenvalue $\lambda$, then 
    \begin{gather*}
        \msf{E}^T(\mbb{R}\Setminus \{\lambda\}) \psi = 0, \\
        \msf{E}^T(\{\lambda\}) \psi = \psi
    \end{gather*}
    hold. Moreover, for any Borel measurable function $f$ such that $f(\lambda)$ is finite, $f(T) \psi = f(\lambda) \psi$ holds.
    }
    Similarly, we can show that $A' = A - \bra A \ket_\theta \mbbm{1} \in \mcal{D}((\gamma_{\msf{E}^A} \gamma_{\msf{E}^A}^{*})^{-1/2})$ and $A' = (\gamma_{\msf{E}^A} \gamma_{\msf{E}^A}^{*})^{-1/2} A'$, since $(\gamma_{\msf{E}^A} \gamma_{\msf{E}^A}^{*})\mbbm{1} = \mbbm{1}$. 
    It follows from Lemma \ref{lem:expllicit formula for tilde JM0} that
    \begin{equation}
        \widetilde{\mcal{J}^{\msf{E}^A}_{0,\theta}}(d_\theta \bra A \ket_\theta) 
        = \| (\gamma_{\msf{E}^A} \gamma_{\msf{E}^A}^{*})^{-1/2} A' \|_{S,\theta}^2
        = \| A' \|_{S,\theta}^2
        = \sigma_\theta(A)^2,
    \end{equation}
    which implies that $\varepsilon_0(A; \theta, \msf{E}^A) = 0$.
\end{proof}

\subsection{Uncertainty relation inequalities}

\begin{theorem}[Watanabe--Sagawa--Ueda uncertainty relation for infinite-dimensional systems]
    Let $\theta \in \Theta$ and let $\msf{M}$ be a POVM on $\mcal{H}$.
    For $A, B \in \mcal{B}(\mcal{H})_{\sa}$ with $\varepsilon (A; \theta, \msf{M}), \varepsilon(B; \theta, \msf{M}) < \infty$, we have
    \begin{equation}
        \varepsilon(A; \theta, \msf{M}) \varepsilon(B; \theta, \msf{M}) \geq \mcal{R}^{\msf{M}}_{\theta}(A,B) + \frac{1}{4} \big| \bra [A,B] \ket_\theta \big|^2,
        \label{eq:uncertainty relation for bounded operators}
    \end{equation}
    where
    \begin{equation}
        \mcal{R}^{\msf{M}}_{\theta}(A,B) := \big|(\JM)^{+}(d_\theta \bra A \ket_\theta, d_\theta \bra B\ket_\theta) - C^S_\theta (A,B)\big|^2.
    \end{equation}
    For self-adjoint $A, B \in \mcal{L}^2_h(\theta)$ with $\varepsilon_0 (A; \theta, \msf{M}), \varepsilon_0(B; \theta, \msf{M}) < \infty$, we have
    \begin{equation}
        \varepsilon_0(A; \theta, \msf{M}) \varepsilon_0(B; \theta, \msf{M}) \geq \mcal{R}^{\msf{M}}_{0,\theta}(A,B) + \frac{1}{4} \big| \bra [A,B] \ket_\theta \big|^2,
        \label{eq:uncertainty relation for unbounded operators}
    \end{equation}
    where
    \begin{equation}
        \mcal{R}^{\msf{M}}_{0,\theta}(A,B) := \big|(\mcal{J}^{\msf{M}}_{0,\theta})^{+}(d_\theta \bra A \ket_\theta, d_\theta \bra B\ket_\theta) - C^S_\theta (A,B)\big|^2.
    \end{equation}
\end{theorem}
\begin{proof}
    Case 1: When $A, B \in \mcal{B}(\mcal{H})_{\sa}$:
    We define a semi-inner product 
    \begin{equation}
        \llangle A, B \rrangle := (\JM)^{+} (d_\theta \bra A \ket_\theta, d_\theta \bra B \ket_\theta) - C_\theta (A,B).
    \end{equation}
    By the Cauchy--Schwarz inequality, we have
    \begin{align*}
        \varepsilon(A; \theta, \msf{M}) \varepsilon(B; \theta, \msf{M}) 
        &= \llangle A, A \rrangle \llangle B, B \rrangle \\
        &\geq \big|\llangle A, B \rrangle\big|^2 \\
        &= \left| (\JM)^{+}(d_\theta \bra A \ket_\theta, d_\theta \bra B \ket_\theta) - \Re C_\theta (A,B) - \im \Im C_\theta (A,B) \right|^2 \\ 
        &= \mcal{R}^{\msf{M}}_{\theta}(A,B) + \left|\frac{1}{2} \bra [A,B] \ket_\theta\right|^2.
        \numberthis
    \end{align*}
    This proves \eqref{eq:uncertainty relation for bounded operators}.

    Case 2: When $A, B \in \mcal{L}^2_h(\theta)$ are self-adjoint: Similarly, the Cauchy--Schwarz inequality for the semi-inner product $\llangle \cdot, \cdot \rrangle_0$ defined by
    \begin{equation}
        \llangle A, B \rrangle_0 := (\mcal{J}^{\msf{M}}_{0,\theta})^{+}(d_\theta \bra A \ket_\theta, d_\theta \bra B \ket_\theta) - C_\theta (A,B)
    \end{equation}
    yields \eqref{eq:uncertainty relation for unbounded operators}.
\end{proof}

\section{Discussion}
\label{sec:discussion}

In this section, we discuss the implications, limitations, and future directions of our extension of the WSU uncertainty relations to infinite-dimensional systems.

\subsection{Physical interpretation of the measurement errors}

In finite-dimensional systems, Watanabe, Sagawa, and Ueda \cite{watanabe2011uncertainty,watanabe2013formulation} defined the measurement error as the difference between the minimum asymptotic variance (multiplied by the sample size $n$) of the estimator of the expectation value $\bra A \ket_\theta$ and the variance $\sigma_\theta(A)^2$.  The former is given by the lower bound of the classical Cram\'{e}r--Rao inequality due to the asymptotic efficiency of the maximum likelihood estimator (MLE) \cite{wald1949MLE,perlman1969strong}.

Following this approach, we defined the measurement errors for infinite-dimensional systems as the difference between the lower bound of the classical Cram\'{e}r--Rao inequality, which is given by the pseudo-inverse of the Fisher information form, and the variance of the operator $A$. Indeed, the measurement errors $\varepsilon(A;\theta,\msf{M})$ and $\varepsilon_0(A;\theta,\msf{M})$ coincide with the original definition in finite-dimensional systems.

However, in infinite-dimensional parameter estimation, the lower bound of the classical Cram\'{e}r--Rao inequality is not always achievable.
Issues such as the non-existence of the unconstrained MLE \cite{geman1982nonparametric}, slower convergence rates than the classical rate \cite{wong1991maximum,birge1993rates,shen1994convergence}, and the need for regularization techniques such as sieves \cite{geman1982nonparametric,wong1991maximum,wong1995probability} must be carefully addressed. 

Nevertheless, the measurement errors defined in this paper remain meaningful as they provide lower bounds for the difference between minimum variance of the estimator and the variance of the operator $A$. In the context of uncertainty relations, demonstrating the impossibility of accurate measurement is crucial, and our definition of measurement errors is well-suited for this purpose. 
For example, if the measurement error $\varepsilon(A;\theta,\msf{M})$ is non-zero, we can conclude that the POVM $\msf{M}$ does not perfectly measure the observable $A$ in the state $\theta$.

\subsection{Improved uncertainty relation bounds}

A notable feature of our results is that the error--error uncertainty relation inequalities \eqref{eq:uncertainty relation for bounded operators} and \eqref{eq:uncertainty relation for unbounded operators} are stronger than the original inequality obtained by WSU. This enhancement arises not from the difference between finite and infinite-dimensional systems, but rather from an improvement in the derivation method. Specifically, our proof utilizes the full Cauchy--Schwarz inequalities for the Hermitian semi-inner products $\llangle \cdot, \cdot \rrangle$ and $\llangle \cdot,\cdot \rrangle_0$, rather than the weaker inequality $\llangle A, A \rrangle \llangle B, B \rrangle \geq (\operatorname{Im} \llangle A, B \rrangle)^2$ used in the original derivation, which was based on the condition $\llangle A + it B, A + it B \rrangle \geq 0$ for all $t \in \mathbb{R}$.

\subsection{Connections to related work}
    A related work is the general formulation of uncertainty relations by Lee and Tsutsui \cite{lee2022local_error-error, LEE2024129962}, which applies to infinite-dimensional quantum systems without requiring specific parameterizations. Their approach recovers the WSU uncertainty relations in finite-dimensional systems under appropriate conditions and includes bounds similar to our $\mathcal{R}^{\msf{M}}_\theta(A,B)$ term. A detailed comparison between their general framework and our specific extension would be valuable for future research.

\subsection{Limitations of the current work and future directions}
A significant limitation of our work is the restriction of the parameter space $\Theta$ to the space of full-rank density operators. This excludes the boundary of the density operator space, particularly pure states, which are of considerable interest in quantum mechanics. The extension of our results to the full space of density operators remains an open problem.

An important challenge for future work is the application of our extended uncertainty relations to concrete infinite-dimensional quantum systems. For example, calculating the measurement errors for observables such as position ${X}$ and momentum ${P}$ operators would provide valuable insights into the practical implications of our theoretical framework. 

Another direction for future research is the extension of the error--disturbance uncertainty relation \cite{watanabe2013formulation} to infinite-dimensional systems. 
The derivation of such relations requires establishing the monotonicity of the quantum Fisher information forms under completely positive trace-preserving (CPTP) maps, which describe general quantum evolution and measurement processes.
While the monotonicity property of the quantum Fisher information operators in finite-dimensional systems is well-established, the extension to infinite-dimensional systems requires careful analysis and is left for future work.

\section*{Acknowledgment}
I wish to thank Professor Shogo Tanimura for his helpful advice. Takemi Nakamura read the Japanese manuscript and gave me constructive suggestions, and Motohide Nakamura spent a lot of time discussing the content with me.
My thanks are also due to Shintaro Minagawa and Naoya Ogawa, who helped me with some of the proofs.

I am grateful to the anonymous reviewer for his or her careful examination of the original manuscript, particularly for identifying inappropriate mathematical formulations and suggesting the use of pseudo-inverse forms, which significantly improved the theoretical framework of this work.

I acknowledge the use of Claude Sonnet 3.7 and Claude Sonnet 4 (Anthropic) for translation assistance and grammar checking during the preparation of this manuscript.

\let\doi\relax

\end{document}